\documentclass[aps,prd, twocolumn,superscriptaddress,nofootinbib,preprintnumbers]{revtex4-1}
\usepackage{amsmath}
\usepackage{graphicx}
\usepackage{subfigure}
\usepackage{amssymb}
\usepackage{xcolor}
\usepackage{multirow}
\usepackage{cancel}
\usepackage{color}
\usepackage{ulem}
\usepackage{listings}
\usepackage{xcolor}
\usepackage{float}
\usepackage{slashed}

\usepackage{tikz}
\usetikzlibrary{arrows,shapes}
\usetikzlibrary{trees}
\usetikzlibrary{matrix,arrows} 				% For commutative diagram
\usetikzlibrary{positioning}				% For "above of=" commands
\usetikzlibrary{calc,through}				% For coordinates
\usetikzlibrary{decorations.pathreplacing}  % For curly braces
\usepackage{pgffor}							% For repeating patterns

\usetikzlibrary{decorations.pathmorphing}	% For Feynman Diagrams
\usetikzlibrary{decorations.markings}
\tikzset{
    vector/.style={decorate, decoration={snake}, draw},
	provector/.style={decorate, decoration={snake,amplitude=2.5pt}, draw},
	antivector/.style={decorate, decoration={snake,amplitude=-2.5pt}, draw},
    fermion/.style={draw=black, postaction={decorate},
        decoration={markings,mark=at position .55 with {\arrow[draw=black]{>}}}},
    fermionbar/.style={draw=black, postaction={decorate},
        decoration={markings,mark=at position .55 with {\arrow[draw=black]{<}}}},
    fermionnoarrow/.style={draw=black},
    gluon/.style={decorate, draw=black,
        decoration={coil,amplitude=4pt, segment length=5pt}},
    scalar/.style={dashed,draw=black, postaction={decorate},
        decoration={markings,mark=at position .55 with {\arrow[draw=black]{>}}}},
    scalarbar/.style={dashed,draw=black, postaction={decorate},
        decoration={markings,mark=at position .55 with {\arrow[draw=black]{<}}}},
    scalarnoarrow/.style={dashed,draw=black},
    electron/.style={draw=black, postaction={decorate},
        decoration={markings,mark=at position .55 with {\arrow[draw=black]{>}}}},
	bigvector/.style={decorate, decoration={snake,amplitude=4pt}, draw},
}

\tikzstyle{block} = [draw, rectangle, 
    minimum height=3em, minimum width=6em]

\usepackage[colorlinks,citecolor=blue]{hyperref}
\usepackage{amsmath}%\ge=\geqslant; \le=\leqslant
\usepackage{wrapfig}

\newcommand{\be}{\begin{equation}}
\newcommand{\ee}{\end{equation}}
\newcommand{\beq}{\begin{equation}}
\newcommand{\eeq}{\end{equation}}
\newcommand{\bea}{\begin{eqnarray}}
\newcommand{\eea}{\end{eqnarray}}
\newcommand{\besp}{\begin{equation}\begin{split}}
\newcommand{\eesp}{\end{split}\end{equation}}

\newcommand{\Eq}[1]{Eq.~(\ref{#1})}
\newcommand{\Dfbd}{\mathord{\buildrel{\lower3pt\hbox{$\scriptscriptstyle\leftrightarrow$}}\over {D}_{\mu}}}
\newcommand{\ave}[1]{\left\langle #1\right\rangle}

\def\mO{\mathcal{O}}

\def\Z{\mathbb{Z}}

\def\0{\textbf{0}}
\def\1{\textbf{1}}
\def\2{\textbf{2}}
\def\3{\textbf{3}}
\def\4{\textbf{4}}
\def\5{\textbf{5}}
\def\6{\textbf{6}}
\def\7{\textbf{7}}
\def\8{\textbf{8}}
\def\9{\textbf{9}}

\def\d{\text{d}}

\begin{document}

\title{Freeze-in of WIMP dark matter}

\author{Xiao-Rui Wang}
\email{xiaorui\_wong@pku.edu.cn}
\affiliation{Department of Physics and State Key Laboratory of Nuclear Physics and Technology,\\ Peking University, Beijing 100871, China}

\author{Ke-Pan Xie}
\email{kpxie@buaa.edu.cn, Corresponding author.}
\affiliation{School of Physics, Beihang University, Beijing 100191, China}

\begin{abstract}

We propose a novel scenario for DM in which weakly interacting massive particles (WIMPs) can freeze-in due to a first-order phase transition (FOPT) in the early Universe. The FOPT dilutes the preexisting DM density to zero and leads to a sudden change in DM mass, preventing WIMPs from re-equilibrating due to their large mass-to-temperature ratio. Following the FOPT, WIMPs are produced via a freeze-in process, even though their interactions are NOT feeble. We demonstrate this concept using a simplified model and then apply it to a realistic model with a delayed electroweak phase transition. Our work presents a promising new direction for the freeze-in mechanism, and also extends the category of WIMP DM.

\end{abstract}

\maketitle

\newpage

\section{Introduction}

Despite its large abundance ($\sim27\%$) in the Universe, the particle origin of dark matter (DM) remains a mystery~\cite{ParticleDataGroup:2022pth}. One of the most promising theoretical paradigms for DM involves assuming that the DM particle $X$ can annihilate into Standard Model (SM) particles via the $2\to2$ scattering
\be\label{2to2}
X\,X\to{\rm SM~SM}.
\ee
Depending on the strength of the portal interaction between the SM and dark sectors, there are two extensively studied scenarios. In the first scenario, \Eq{2to2} is in thermal equilibrium in the early Universe, causing DM particles to follow the equilibrium distribution until the temperature drops to $\sim1/25$ of the DM mass, at which point the annihilation process decouples and a fixed DM relic abundance remains. This process is known as the freeze-out of weakly interacting massive particles (WIMPs)~\cite{Chiu:1966kg,Lee:1977ua,Roszkowski:2017nbc}, which has been the most popular explanation for particle DM. In the second scenario, the initial density of DM is negligibly small, and the interactions are so feeble that DM particles can never reach thermal equilibrium. As a result, DM accumulates via the inverse process of \Eq{2to2}, leading to the freeze-in of feebly interacting massive particles (FIMPs)~\cite{Hall:2009bx,McDonald:2001vt,Bernal:2017kxu}.

The reaction \Eq{2to2} can realize two opposite scenarios, namely WIMP freeze-out and FIMP freeze-in. In this work, we propose a novel scenario based on the same reaction, which is the {\it freeze-in} of the WIMPs. By ``WIMPs,'' we mean that the portal interactions are not feeble. Therefore, in the conventional thermal history of the Universe, DM particles inevitably thermalize and freeze-out. However, we suggest that freeze-in of WIMPs can happen if the Universe experiences a supercooled first-order phase transition (FOPT). A FOPT is the transition of the Universe from a metastable false vacuum to a stable true vacuum via bubble nucleation and expansion~\cite{Hindmarsh:2020hop}, and its usage is two-fold:
\begin{enumerate}
\item A supercooled FOPT releases a huge amount of entropy, which dilutes the preexisting DM density to a negligible level.
\item The WIMPs gain mass from the FOPT, such that after the transition the DM particles have a huge mass-to-temperature ratio and hence an exponentially suppressed Boltzmann factor, which prevents them from thermalizing.
\end{enumerate}
Therefore, after the FOPT, DM will be accumulatively produced via the inverse process of \Eq{2to2}, which is a typical freeze-in scenario, but it applies to weak or moderate couplings, rather than feeble ones as seen in traditional FIMP freeze-in.

Our work introduces a novel scenario for DM based on the simple $2\to2$ annihilation and represents a third possible scenario in addition to the traditional WIMP freeze-out and FIMP freeze-in. As will be demonstrated, this scenario shares the common features from the conventional freeze-in~\cite{Hall:2009bx}, such as the DM behavior is determined by physics at and below the scale of the DM mass but independent of higher scale such as the inflationary models; and the relic abundance is positively correlated with the coupling strength, etc.

\section{Freeze-in of WIMPs}

We illustrate the idea using a simplified model with a scalar DM candidate $X$ that interacts with a massless thermal bath scalar $B$ via the quartic coupling $\lambda X^\dagger X B^\dagger B$. In the radiation era, the Boltzmann equation governing the evolution of $X$ is
\be\label{Boltzmann}
\frac{\d Y_X}{\d z}=-\sqrt{\frac{\pi g_{*,s}^2}{45g_*}}\frac{M_{\rm Pl}m_X}{z^2}\ave{\sigma v_{\rm rel}}\left(Y_X^2-Y_{\rm eq}^2\right),
\ee
where $Y_X=n_X/s$ is the yield of $X$ with $s$ being the entropy density, $z=m_X/T$, $M_{\rm Pl}=1.22\times10^{19}$ GeV is the Planck scale, $g_{*,s}$ and $g_*$ are the numbers the relativistic degrees of freedom for entropy and energy, respectively, $Y_{\rm eq}=45z^2K_2(z)/(4\pi^4g_{*,s})$ is the equilibrium yield of $X$,
\be
\ave{\sigma v_{\rm rel}}=\frac{\lambda^2}{32\pi m_X^2}\left(\frac{K_1(z)}{K_2(z)}\right)^2
\ee
is the thermal average of the annihilation cross section of $XX^\dagger\to BB^\dagger$ multiplying the relative velocity $v_{\rm rel}$ under the Maxwell-Boltzmann distribution, and $K_i(z)$ is the $i$-th modified Bessel function.

In the conventional thermal history, $z$ starts from $\sim0$ at the end of the inflationary reheating epoch and evolves to $\gg1$ to the current Universe. If $\lambda$ is sufficient to keep $X$ in equilibrium for $z\ll1$, then \Eq{Boltzmann} realizes the WIMP freeze-out scenario that $\Omega_Xh^2\sim0.1\,(0.5/\lambda)^2(m_X/{\rm TeV})^2$, which implies an upper limit of $\sim100$ TeV for the DM mass due to the unitarity bound of $\lambda$, known as the GK bound~\cite{Griest:1989wd}. On the other hand, for feeble $\lambda$, \Eq{Boltzmann} explains DM with a FIMP freeze-in scenario that has $\Omega_Xh^2\sim0.1\,[\lambda/(2.5\times10^{-11})]^2$, independent of the DM mass.

\begin{figure}
\centering
\subfigure{
\includegraphics[scale=0.45]{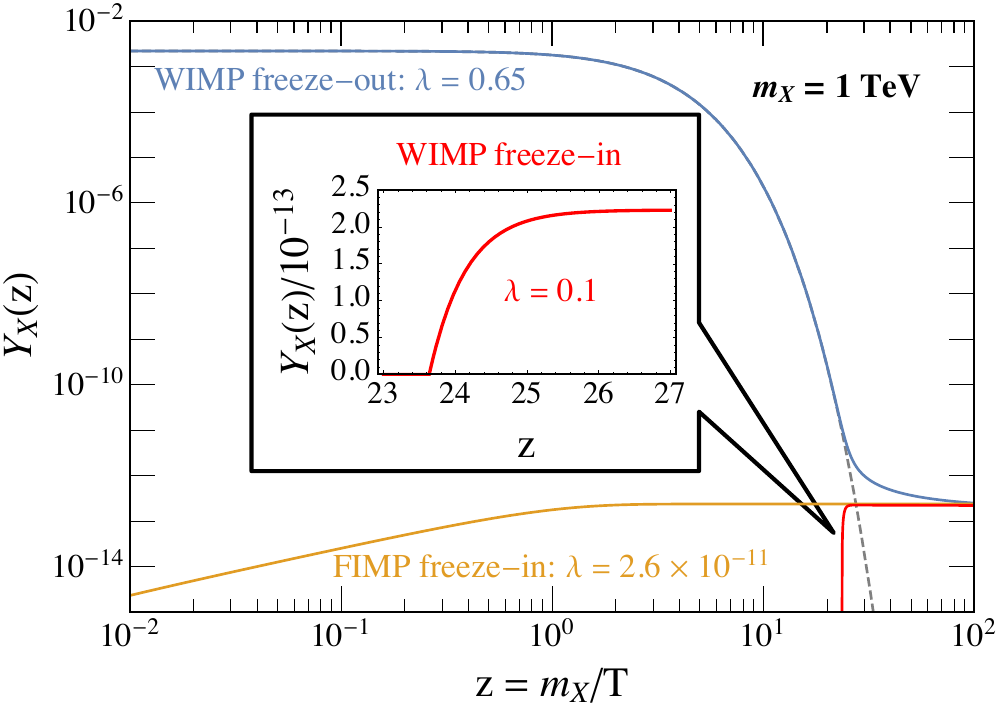}}
\caption{The three DM scenarios realized by the Boltzmann equation (\ref{Boltzmann}) for $m_X=1$ TeV with different $\lambda$. The blue, orange and red lines are WIMP freeze-out ($\lambda\approx0.65$), FIMP freeze-in ($\lambda\approx2.6\times10^{-11}$) and WIMP freeze-in ($\lambda=0.1$ and $z_2\approx23.6$), respectively. The equilibrium distribution is plotted in gray dashed line.}\label{fig:sketch}
\end{figure}

In our WIMP freeze-in model, there exists a discontinuity in the evolution of $z$ during the thermal history. Prior to the FOPT, $X$ is massless, thus $z\equiv0$ and $Y_1\approx(2/\pi^2)/(2\pi^2g_{*,s}/45)\sim\mathcal{O}(10^{-3})$. Following the transition, however, the DM mass undergoes a sudden change to $m_X\gg T_2$, where $T_2$ denotes the temperature after the FOPT. We assume a supercooled FOPT such that $T_2\gg T_1$, the temperature at which the FOPT begins. This leads to an enormous increase in entropy density by a factor of $(T_2/T_1)^3$. Consequently, the evolution of \Eq{Boltzmann} begins at $z_2=m_X/T_2\gg1$, with an initial condition $Y_X(z_2)\approx (T_1/T_2)^3Y_1\sim0$. Freeze-in then occurs via $BB^\dagger\to XX^\dagger$, and the yield can be approximately solved from \Eq{Boltzmann}
\be\label{FI_estimate}
Y_{\infty}\approx\frac{135\sqrt{5}\lambda^2M_{\rm Pl}}{4096\pi^{15/2}g_{*,s}\sqrt{g_*}m_X}(1+2z_2)e^{-2z_2},
\ee
and hence $\Omega_Xh^2\sim0.1\times\left[\lambda e^{-z_2}/(3.5\times10^{-11})\right]^2(1+2z_2)$,
where we use $g_*\approx g_{*,s}\approx106.75$. Even if $\lambda$ is NOT feeble, this scenario can still produce a correct DM relic abundance via a large enough $z_2$ provided by a strong FOPT, and that is the crucial point for the WIMP freeze-in. For a given $z_2$, $\Omega_Xh^2$ is proportional to $\lambda^2$ but irrelevant to $m_X$; this implicit dependence allows for superheavy DM exceeding the GK bound. Reducing the FOPT strength decreases $z_2$ and requires a corresponding decrease in $\lambda$ to maintain $\lambda e^{-z_2}\sim10^{-11}$, as indicated by \Eq{FI_estimate}. In the no FOPT limit ($z_2\to0$), $\lambda\to\mO(10^{-11})$, transitioning to conventional FIMP freeze-in.

Fig.~\ref{fig:sketch} illustrates the three DM scenarios for $m_X=1$ TeV and different $\lambda$ values, all give the correct DM abundance $\Omega_{\rm DM}h^2=0.12$~\cite{ParticleDataGroup:2022pth}. The gray dashed line is the equilibrium $X$ distribution for reference. For $\lambda\approx0.65$, the WIMP freeze-out is realized in the blue line; while for $\lambda\approx2.6\times10^{-11}$, the FIMP freeze-in is given in the orange line. Our WIMP freeze-in scenario is described by the red line, corresponding to $\lambda=0.1$ and $z_2\approx23.6$: the DM yield starts from zero and increases rapidly to a fixed value at around $z\sim25$. By decreasing $z_2$ and adjusting $\lambda$ accordingly to maintain the correct DM yield, the red line gradually shifts leftwards and eventually aligns with the orange line in the no FOPT limit, i.e. $z_2\to0$. This again shows the connection between our scenario and the conventional freeze-in.

\section{Model building and phenomenology}

The necessary supercooled FOPT could be realized in a classically conformal (CC) model, whose scalar potential has no quadratic mass term at tree level but a Coleman-Weinberg (CW) potential~\cite{Coleman:1973jx} is generated at one-loop level~\cite{Iso:2009ss,Iso:2009nw,Das:2015nwk}. It is well-known that such models can exhibit supercooled FOPTs~\cite{Konstandin:2011dr,Jinno:2016knw,Iso:2017uuu,Sagunski:2023ynd,Marzo:2018nov,Bian:2019szo,Ellis:2019oqb,Ellis:2020nnr,Jung:2021vap,Huang:2022vkf,Khoze:2022nyt}. The minimal setup is that $\phi=\sqrt{2}\,{\rm Re}[B]$ is the FOPT scalar field, and its potential can be parametrized as
\be\label{CW}
V_1(\phi)=V_\Lambda+\frac{\lambda_B^2}{64\pi^2}\phi^4\left(\log\frac{\phi}{w}-\frac14\right),
\ee
where $V_\Lambda=\lambda_B^2w^4/(256\pi^2)$ is the vacuum energy, and $\lambda_B$ receives contributions from all particles coupling to $B$. The potential yields a vacuum expectation value (VEV) $\ave{\phi}=w$, which breaks the CC symmetry spontaneously and provides a mass $m_X^2=\lambda w^2/2$ to the DM.

\Eq{CW} can trigger a supercooled FOPT from $\phi=0$ to $\phi\approx w$ at a temperature of $T_1\ll w$, releasing a significant amount of vacuum energy $V_\Lambda$ and reheating the Universe to $T_2\approx T_\Lambda$, where $\pi^2g_*T_\Lambda^4/30=V_\Lambda$. The dilution condition requires $ (T_1/T_2)^3Y_1\ll Y_{\rm DM}\approx0.8~{\rm eV}/m_X$, which is translated to $T_2/T_1\gg 2000\times(m_X/{\rm TeV})^{1/3}$. By substituting these equations into \Eq{FI_estimate} and requiring the correct DM abundance, we obtain
\be\label{numerical_estimate}
\lambda_B\approx0.189\,\lambda^{0.881},
\ee
which provides a relation between the DM coupling $\lambda$ and the potential coefficient $\lambda_B$, and can be treated as a guide for model building. For example, if we would like to build a model with minimal particle content, then $\lambda_B\sim\lambda$ and \Eq{numerical_estimate} yields $\lambda\sim10^{-6}$, which is the expected coupling strength for WIMP freeze-in. If one instead favors a coupling at scale of electroweak (EW) gauge coupling, e.g. $\lambda\sim g_2^2\approx0.4$, then \Eq{numerical_estimate} estimates $\lambda_B\approx0.1<\lambda$, implying additional fermionic degrees of freedom coupled to $B$, which provide negative contributions to $\lambda_B$.

Our scenario can be probed through the direct~\cite{Schumann:2019eaa}, indirect~\cite{Gaskins:2016cha} and collider~\cite{Boveia:2018yeb} searches as in the traditional WIMP scenario. In addition, the stochastic gravitational wave (GW) background is another approach to probe this scenario since supercooled FOPTs generate strong GWs~\cite{Ellis:2019oqb,Ellis:2020nnr}. The correlation between GW and WIMP signals could efficiently probe the scenario.

\section{A realistic model}

Extend the SM with three gauge singlets: one real scalar $\phi$, one complex DM scalar $X$, and one Dirac fermion $\psi$. The tree level CC potential is $V(H,\phi)=\lambda_h|H|^4+\lambda_x|X|^4+\frac{\lambda_\phi}{4}\phi^4
+\lambda_{hx}|H|^2|X|^2+\frac{\lambda_{h\phi}}{2}\phi^2|H|^2+\frac{\lambda_{\phi x}}{2}\phi^2|X|^2$~\cite{Haruna:2019zeu,Hamada:2020wjh,Hamada:2021jls,Kawana:2022fum}, where $H=(G^+,(h+iG^0)/\sqrt{2})^T$ is the SM Higgs doublet. $\psi$ couples to other particles via $-(y_\psi/\sqrt2)\phi\bar\psi\psi-y_\nu\bar\ell_L\tilde H\psi$, with $\ell_L$ the SM lepton doublet. The one-loop CW potential is replacing $\lambda_B^2$ with $(\lambda_{\phi x}^2-2y_\psi^4)$ in \Eq{CW}. This leads to $\ave{\phi}=w$ and the breaking of CC symmetry, which also triggers the EW symmetry breaking via $\lambda_{h\phi}\approx-m_h^2/w^2$, leading to $\ave{h}=v_{\rm EW}\approx246$ GeV~\cite{Iso:2009ss}. The particle spectrum includes one Dirac fermion $\psi$ with $m_\psi=y_\psi w/\sqrt2$ and three scalar bosons with $m_\phi\approx\sqrt{\lambda_{\phi x}^2-2y_\psi^4}w/(4\pi)$, $m_X=\sqrt{(\lambda_{\phi x}w^2+\lambda_{hx}v_{\rm EW}^2)/2}$ and $m_h\approx125$ GeV. An unbroken $\Z_2$ symmetry ensures $X$'s stability.

At finite temperature, the scalar potential becomes $V_T(\phi)\approx V_1(\phi)+(\lambda_{\phi x}+y_\psi^2)T^2\phi^2/24$. When $T\gg w$, the Universe stays in the EW-preserving vacuum $(\phi,h)=(0,0)$. As $T$ drops, the potential develops another local minimum at $\phi\sim w$, which eventually becomes the true vacuum, i.e. the global minimum. However, when $(\lambda_{\phi x}+y_\psi^2)$ is small, the tunneling probability is too small to trigger the FOPT from $\phi=0$ to $\phi\sim w$, and the Universe is trapped in the false vacuum.

\begin{figure}[b]
\centering
\subfigure{
\includegraphics[scale=0.45]{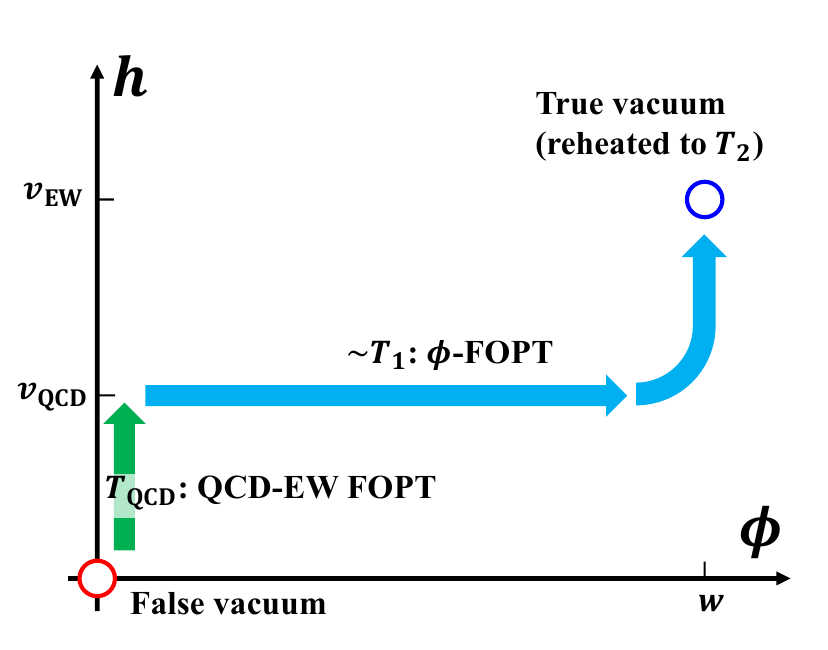}}
\caption{The thermal history of the realistic model in the field space. The Universe is trapped in the origin down to $T_{\rm QCD}\approx85$ MeV, when the QCD-EW FOPT occurs. Then at $\sim T_1$, the $\phi$-FOPT happens and the Universe rolls down to the true vacuum, reheating the Universe to $T_2$.}\label{fig:field_space}
\end{figure}

If the Universe stays in the $\phi$-$h$ space origin until $T_{\rm QCD}\approx85$ MeV, then the QCD chiral phase transition occurs via a FOPT~\cite{Braun:2006jd}, as there are 6 massless quarks in the plasma~\cite{Pisarski:1983ms}. The top quark condensate then generates a Higgs VEV $\ave{h}=v_{\rm QCD}=(y_t\ave{\bar tt}/\sqrt{2}\lambda_h)^{1/3}$~\cite{Iso:2017uuu,Sagunski:2023ynd}, triggering a QCD-EW FOPT from $(\phi,h)=(0,0)$ to $(0,v_{\rm QCD})$. Then the potential becomes $V_T(\phi)\approx V_1(\phi)+[(\lambda_{\phi x}+y_\psi^2)T^2+6\lambda_{h\phi}v_{\rm QCD}^2]\phi^2/24$, still trapping the $\phi$ field in its origin until the the Universe cools to
\be
T_1=v_{\rm QCD}\sqrt{\frac{-6\lambda_{h\phi}}{\lambda_{\phi x}+y_\psi^2}}\approx v_{\rm QCD}\frac{m_h}{w}\sqrt{\frac{6}{\lambda_{\phi x}+y_\psi^2}},
\ee
and the $\phi$-quadratic term vanishes. At $\sim T_1$, the Universe tunnels along the $\phi$ direction and rolls down to the true vacuum $(\phi,h)\approx(w,v_{\rm EW})$, leading to a $\phi$-FOPT, reheating the Universe to $T_2$. The thermal history in $\phi$-$h$ field space is sketched in Fig.~\ref{fig:field_space}, and the full expression of the potential is given in Appendix~\ref{app:potential}.

After the QCD-EW FOPT, $X$'s mass increases from 0 to $m'_X=v_{\rm QCD}\sqrt{\lambda_{hx}/2}$. Then the $\phi$-FOPT enhances the $X$ mass to $m_X\approx\sqrt{(\lambda_{\phi x}w^2+\lambda_{hx}v_{\rm EW}^2)/2}$, generating $z_2=m_X/T_2$ and diluting $X$ yield to
\be\label{dilution}
Y_X(z_2)=Y_{\rm eq}(m'_X/T_1)\frac{T_2}{T_\Lambda}\left(\frac{T_1}{T_\Lambda}\right)^3.
\ee
$T_2=T_\Lambda\min\{1,\Gamma/H\}^{1/2}$, with $\Gamma=\Gamma_h\sin^2\theta+\Gamma_\phi\cos^2\theta$, where $\Gamma_{h,\phi}$ are the decay widths of $h$ and $\phi$, respectively, and $\theta\approx-v_{\rm EW}/w$ is the mixing angle~\cite{Hambye:2018qjv}. In the parameter space of interest, reheating is prompt ($\Gamma\gg H$) and hence $T_2=T_\Lambda$. We find $T_2\sim\mO(100)$ GeV, which might restore the EW symmetry; however, $T_2/w\lesssim10^{-2}$ and hence $\ave{\phi}$ is not affected.

\begin{figure}
\centering
\subfigure{
\includegraphics[scale=0.4]{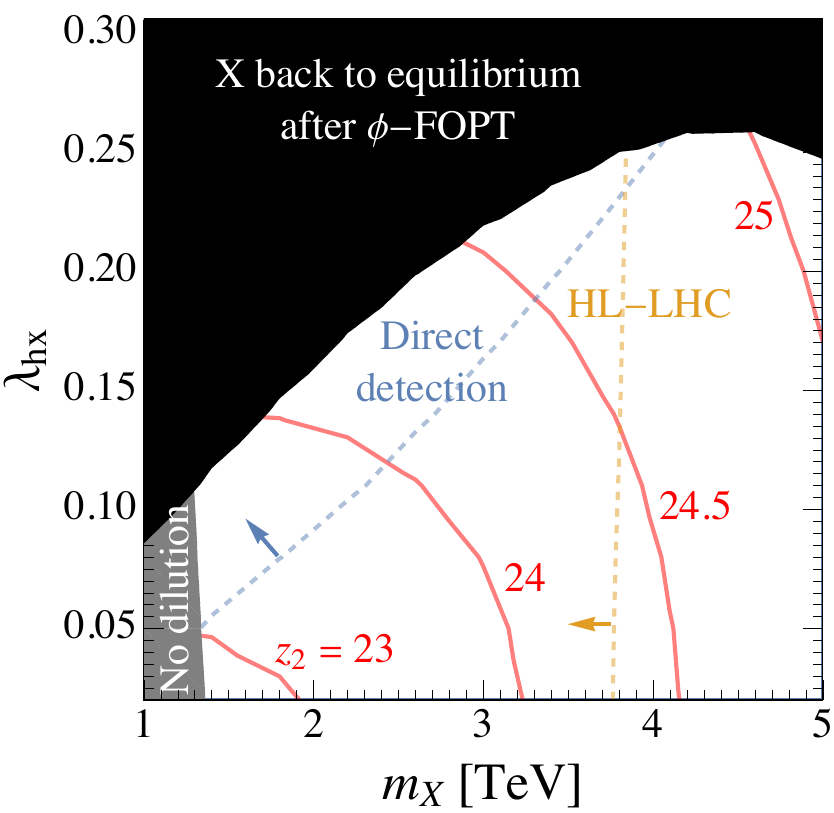}}
\caption{The parameter space that realizes the WIMP freeze-in scenario. The red contours are $z_2$ that give the correct DM density. The gray-shaded region cannot dilute the preexisting DM to a negligible level, while the black-shaded region cannot prevent the DM from thermalization. The parameter space that can be probed by direct detection and Higgs exotic decay is plotted with the blue and orange lines, respectively.}\label{fig:scan}
\end{figure}

After the FOPT, $X$ is produced via $\phi\phi, hh\to XX^\dagger$. Taking $v_{\rm QCD}=100$ MeV and $w=10$ TeV as a benchmark, given a set of $(m_X,\lambda_{hx})$, we derive $\lambda_{\phi x}$ and the $y_\psi$ required for the correct DM relic abundance, and present the results in Fig.~\ref{fig:scan}. We have incorporated thermal effects on scalar masses, and checked $\phi\phi\to XX^\dagger$ dominates in the bottom-right area, whereas $hh\to XX^\dagger$ dominates in the top-left area.

Large $\lambda_{hx}$ leads to $X$ particles returning to equilibrium and undergoing normal freeze-out after the FOPT, as covered by the black region; while the region fails to satisfy the dilution condition of $Y_X(z_2)\ll Y_{\rm DM}$ is covered by gray. The white region enables the realization of the WIMP freeze-in scenario, with corresponding values of $z_2$ indicated by red contours. The portal couplings $\lambda_{hx,\phi x}\sim0.1$, consistent with \Eq{numerical_estimate}.

The spin-independent $X$-nucleon elastic scattering cross section is $\sigma_{\rm SI}\sim10^{-48}~{\rm cm}^2$, challenging in direct detection. Nonetheless, a considerable fraction of the parameter space has a $\sigma_{\rm SI}$ larger than the neutrino floor (shown as the blue dashed line) and hence might be probed by future experiments~\cite{LZ:2018qzl}. The light $\phi$ boson leads to Higgs exotic decay, which can be probed at the HL-LHC by projected reach ${\rm Br}(h\to\phi\phi)\approx4\%$~\cite{deBlas:2019rxi}, plotted as the dashed orange line in Fig.~\ref{fig:scan}. Besides, the ratio of $\phi$-FOPT latent heat to the radiation energy is typical $\alpha\gtrsim10^{14}$, the GWs are very strong and mainly from bubble collisions~\cite{Ellis:2019oqb,Ellis:2020nnr}. Taking the ratio of the Hubble time scale to FOPT duration $\beta/H_*=100$ as a benchmark, we estimate the GW spectra~\cite{Caprini:2015zlo,Caprini:2019egz}. The GWs peak at $f\sim10^{-3}$ Hz with $\Omega_{\rm gw}h^2\sim10^{-10}$, within the sensitive region of a few near-future space-based interferometers, including LISA~\cite{LISA:2017pwj}, TianQin~\cite{TianQin:2015yph,TianQin:2020hid}, Taiji~\cite{Hu:2017mde,Ruan:2018tsw}, BBO~\cite{Crowder:2005nr} and DECIGO~\cite{Kawamura:2011zz}. The projected reach of the 1-year operation of LISA, TianQin and Taiji can cover the parameter space in Fig.~\ref{fig:scan}.

\section{Conclusion}

We propose a novel DM scenario based on the simple $2\to2$ annihilation process, showing that WIMP freeze-in is viable with the assistance of a FOPT. Although we illustrate the idea with scalar DM, similar discussion applies to fermion DM as well. This work focuses on the extreme case where $X$ is initially massless and gains mass after the FOPT. This requires a large (but achievable) dilution factor; but it also extends to cases where $X$ particles have a pre-FOPT mass of $m_{X1}$ and experience a mass jump to $m_{X2}\gg T_2$ after the FOPT. In such instances, the dilution condition is much relaxed. Our mechanism generally applies to many new physics models with background supercooled FOPTs.

It is known that DM evolution is affected by non-standard thermal history such as early matter era~\cite{Cosme:2020mck,Cirelli:2018iax,Chaudhuri:2021bmn}, second-order PTs~\cite{Heurtier:2019beu,Hashino:2021dvx,Elor:2021swj}, non-thermal production after inflationary reheating~\cite{Kuzmin:1997jua,Chung:1998rq,Giudice:2000ex,Nee:2022pxx}, and FOPTs can leave great impacts on WIMP freeze-out or FIMP freeze-in via the change of particle masses~\cite{Cohen:2008nb,Baker:2017zwx,Bian:2018mkl,Bian:2018bxr}. Besides, FOPTs alter the decay of DM~\cite{Baker:2016xzo,Kobakhidze:2017ini,Baker:2018vos,DiBari:2020bvn,Kobakhidze:2019tts}, produce DM non-thermally~\cite{Falkowski:2012fb,Baldes:2020kam,Azatov:2021ifm,Baldes:2022oev}, filter the DM particles~\cite{Baker:2019ndr,Chway:2019kft,Chao:2020adk}, dilute the DM density~\cite{Hambye:2018qjv,Baldes:2021aph}, or form macroscopic DM including primordial black holes~\cite{Krylov:2013qe,Huang:2017kzu,Bai:2018vik,Bai:2018dxf,Atreya:2014sca,Hong:2020est,Baker:2021nyl,Kawana:2021tde,Liu:2021svg,Baker:2021sno,Hashino:2021qoq,Huang:2022him,Bai:2022kxq,Kawana:2022lba,He:2022amv}. The WIMP freeze-in proposed in this work provides a new connection between FOPTs and DM, opening up a third possibility for realizing DM besides the traditional WIMP freeze-out and FIMP freeze-in mechanisms, allowing for WIMPs with mass beyond the GK bound, and it can be tested by combining the WIMP and GW experiments.

Finally, we clarify the novelty of our work in comparison to several related scenarios. One of the key ingredients of our scenario is the dilution of preexisting DM density, which can be induced by entropy injections from processes other than FOPTs, such as the late-time decay of a heavy particle dominating the Universe's energy. However, such a scenario relies on an exceedingly long lifetime for the heavy particle, which in turn necessitates an extremely weak interaction strength that gives rise to another FIMP, see the Appendix~\ref{app:decay}. In contrast, our proposed supercooled FOPT scenario allows for entropy injection with a moderate coupling, as demonstrated by the realistic model with $\lambda_{\phi x}\sim\mO(0.1)$. Hence, the FOPT scenario emerges as a preferable mechanism since it achieves freeze-in without introducing additional FIMPs.

Given the similarities between FOPT reheating and inflationary reheating, one may wonder if a similar WIMP freeze-in mechanism occurs when the DM mass exceeds the reheat temperature from inflation. Here, we clarify the crucial distinction between these two scenarios. In the inflation-induced scenario, the decay width of the inflaton is typically much smaller than the inflationary Hubble scale $H_I$. Therefore, reheating is a slow process, during which the maximal temperature $T_{\rm max}$ is usually significantly higher than the final reheat temperature $T_{\rm rh}$. Consequently, DM particles with masses up to $2000\,T_{\rm rh}$ can be abundantly produced without exponential suppression~\cite{Kuzmin:1997jua,Chung:1998rq,Giudice:2000ex,Nee:2022pxx}. This requires either a feeble coupling or a superheavy DM particle whose mass is related to $H_I$ to reduce the relic abundance to current observed value, see the Appendix~\ref{app:inflation}. In contrast, in the FOPT-induced scenario, the scalar decay width easily surpasses the Hubble constant at FOPT, leading to instantaneous reheating, and the temperature increases monotonically from $T_1$ to $T_2$. Subsequently, DM undergos freeze-in, resulting in $\Omega_Xh^2\propto e^{-2m_X/T_2}$, and the observed DM can be explained by $m_X\sim20\,T_2$. The novelty of our scenario lies in its ability to naturally achieve the desired exponential suppression factor, independent of the cosmic history before the FOPT.

\section*{Acknowledgement}

We thank Iason Baldes, Huai-Ke Guo, Chengcheng Han, Kiyoharu Kawana, Tong Li, Wei Liu, Kengo Shimada and Tao Xu for the very useful discussions.

\appendix

\section{The full expression of scalar potential in the realistic model}\label{app:potential}

The one-loop CW potential is
\be\label{CW1}
V_1(\phi)=V_\Lambda+\frac{\lambda_{\phi x}^2-2y_\psi^4}{64\pi^2}\phi^4\left(\log\frac{\phi}{w}-\frac14\right),
\ee
where $V_\Lambda=(\lambda_{\phi x}^2-2y_\psi^4)w^4/(256\pi^2)$. The contribution of Higgs field to $V_1(\phi)$ is negligible due to the small $|\lambda_{h\phi}|\approx m_h^2/w^2$. In the early Universe, the scalar potential receives thermal corrections and becomes~\cite{Kawana:2022fum}
\begin{widetext}
\be
V_T(\phi)=V_1(\phi)+\frac{2T^4}{2\pi^2}J_B\left(\frac{\lambda_{\phi x}\phi^2}{2T^2}\right)+\frac{4T^4}{2\pi^2}J_F\left(\frac{y_\psi^2\phi^2}{2T^2}\right)
-\frac{2T}{12\pi}\left(\frac{\lambda_{\phi x}}{2}\right)^{3/2}\left[\left(\phi^2+\frac{T^2}{12}\right)^{3/2}-\phi^3\right],
\ee
\end{widetext}
where the bosonic and fermionic thermal integrals are
\be
J_{B/F}(y)=\pm\int_0^\infty x^2{\rm d}x\ln\left(1\mp e^{-\sqrt{x^2+y}}\right).
\ee
For $y\lesssim2$, the high-temperature expansions are
\be\begin{split}
J_B(y)\approx&~-\frac{\pi^4}{45}+\frac{\pi^2}{12}y-\frac{\pi}{6}y^{3/2}-\frac{y^2}{32}\log\frac{y}{a_B},\\
J_F(y)\approx&~-\frac{7\pi^4}{360}+\frac{\pi^2}{24}y+\frac{y^2}{32}\log\frac{y}{a_F},
\end{split}\ee
where $a_B=16a_F$ and $a_F=\pi^2e^{1.5-2\gamma_E}$ with $\gamma_E\approx0.577$ the Euler's constant.

\section{The DM dilution induced by heavy particle decay}\label{app:decay}

The decay of a heavy particle dominating the Universe's energy density can inject entropy into the plasma, leading to the dilution of the DM yield $Y_X(z)$ at a finite $z$. Consequently, this provides the necessary initial condition for WIMP freeze-in without a FOPT. Here, we present a concise quantitative estimation to this scenario. Let us consider a model consisting of a DM candidate $X$ with mass $m_X$ and a heavier particle $N$ with mass $m_N > m_X$. No FOPT occurs during the thermal history, and hence there is no mass jump for the DM particle. The long lifetime of $N$ is ensured by its extremely small decay width, denoted as $\Gamma_N$. Prior to the decay of $N$, both $X$ and $N$ attain frozen yields
\be\label{YXYN}
Y_X^{\rm fo} \propto \frac{m_X}{\alpha_X^2}, \quad Y_N^{\rm fo}\propto \frac{m_N}{\alpha_N^2},
\ee
through the standard freeze-out mechanism, where $\alpha_X$ and $\alpha_N$ represent the finite structure constants corresponding to the $2\to2$ annihilation processes of $X$ and $N$ respectively. At late-time, $N$ dominates the energy of the Universe and decays to light particles, reheating the Universe and diluting the $X$ yield to~\cite{kolb2018early}
\be
Y_X^{\rm fo}\to Y_X(z_2)\approx Y_X^{\rm fo}\times\frac{1}{1.83\ave{g_*^{1/3}}^{3/4}}\frac{1}{Y_N^{\rm fo}}\frac{\sqrt{M_{\rm Pl}\Gamma_N}}{m_N}.
\ee

In order to achieve the WIMP freeze-in scenario, the diluted yield should be significantly smaller than the current observed DM yield. Specifically, we require $Y_X(z_2) \ll Y_{\text{DM}} \approx 0.8~\text{eV}/m_X$, where $Y_{\text{DM}}$ represents the observed DM yield. This condition translates into a constraint on the decay width $\Gamma_N \approx y_N^2m_N/(8\pi)$ and, consequently, the coupling strength $y_N$,
\be
y_N\ll2\times10^{-19}\left(\frac{g_*}{100}\right)^{1/4}\left(\frac{\rm TeV}{m_X}\right)^{1/2}\left(\frac{m_N}{m_X}\right)^{3/2}\left(\frac{\alpha_X}{\alpha_N}\right)^2,
\ee
where \Eq{YXYN} has been used. If we assume that the mass $m_N$ of the heavy particle is not significantly different from the scale of $m_X$ to avoid UV sensitivity, we observe that achieving the WIMP freeze-in mechanism in the heavy-particle-decay scenario necessitates a small coupling $y_N$ to guarantee an adequately long lifetime for the heavy particle $N$. If $\alpha_X\gg\alpha_N$, the bound on $y_N$ can be relaxed, however this introduces another extremely small coupling $\alpha_N$. In either case, we always have a heavy particle with feeble interactions, which is the feature of this heavy-particle-decay scenario.

The novelty of our FOPT-induced WIMP freeze-in scenario is that it does not need any FIMPs, and the dilution naturally occurs through entropy injection from a supercooled FOPT. In our realistic model, for instance, a value of $\lambda_{\phi x}\sim \mO(0.1)$ allows for a strong FOPT that effectively dilutes the preexisting $X$, even if it is initially massless and therefore abundant.

\section{The inflation-induced WIMP freeze-in scenario}\label{app:inflation}

If the DM mass exceeds the inflationary reheating temperature, the DM may be produced through freeze-in with a relic abundance exponentially suppressed. This represents a viable WIMP freeze-in scenario under some conditions, but it differs fundamentally from our FOPT-induced mechanism due to the difference between Hubble constants and the resulting dynamics during inflationary reheating or FOPT reheating. We provide an outline of the two cases below.

We first acknowledge the similarity between the two scenarios. In both cases, a scalar field transfers its energy to the plasma, reheating the Universe and diluting the preexisting DM density to establish suitable initial conditions for freeze-in. In the inflation-induced scenario, the inflaton field $\varphi$ undergoes coherent oscillations around the minimum at the end of inflation and subsequently decays, leading to the reheating of the Universe to a temperature $T_{\rm rh}$ that marks the beginning of the radiation era. On the other hand, in the FOPT-induced scenario, the scalar field responsible for the FOPT, denoted as $\phi$, tunnels across the potential barrier at $T_1$ and rolls down to the true vacuum, resulting in the reheating of the Universe to a temperature $T_2$. The crucial distinction between the two scenarios arises from the dynamics of reheating.

In the inflation-induced scenario, the decay width $\Gamma_\varphi$ of the inflaton field is typically much smaller than the inflationary Hubble constant $H_I$. For instance, for a reheating temperature $T_{\rm rh}\sim\sqrt{\Gamma_\varphi M_{\rm Pl}}\sim 10^9$ GeV, we find $\Gamma_\varphi\sim10^{-1}$ GeV, significantly smaller than the typical value of $H_I\sim10^{13}$ GeV. This results in a slow reheating process where the inflaton field $\varphi$ undergoes multiple oscillations around the potential minimum. During this period, the temperature of the Universe first reaches a maximum value $T_{\rm max}\sim(H_IM_{\rm Pl})^{1/4}T_{\rm rh}^{1/2}$ and then decreases as $T\propto a^{-3/8}$, with $a$ denoting the FLRW scale factor. Notably, the maximal temperature $T_{\rm max}\gg T_{\rm rh}$. Consequently, a huge amount of $X$ can be produced during reheating, and the resultant relic abundance is~\cite{Chung:1998rq,Giudice:2000ex}
\be
\Omega_Xh^2\approx 10^{14}\times m_X^2\ave{\sigma v_{\rm rel}}\left(\frac{g_*}{200}\right)^{-3/2}\left(\frac{20}{m_X/T_{\rm rh}}\right)^7,
\ee
for $m_X$ up to $\sim2000\,T_{\rm rh}$, without the expected exponential suppression factor of $e^{-2m_X/T_{\rm rh}}$. To match $\Omega_Xh^2$ with the current $\Omega_{\rm DM}h^2\sim0.1$, one then needs an extremely small coupling responsible for the $XX^\dagger$ annihilation cross section $\ave{\sigma v_{\rm rel}}$. For superheavy masses $m_X\gg T_{\rm max}$, the production rate is exponentially suppressed by a factor of $e^{-2m_X/T_{\rm max}}$. However, this introduces a form of UV-sensitivity since $T_{\rm max}$ is connected to the inflationary Hubble constant $H_I$.

In our FOPT-induced WIMP freeze-in scenario, the situation is qualitatively different. The decay width $\Gamma_\phi$ of the FOPT scalar is typically much larger than the Hubble constant during the FOPT, $H_\Lambda\sim\sqrt{V_\Lambda}/M_{\rm Pl}$, where $V_\Lambda=\lambda_B^2w^4/(256\pi^2)$ is the vacuum energy. In our realistic model discussed in the article, with the benchmark $w=10$ TeV and $\lambda_B\sim\mathcal{O}(1)$, we have $H_\Lambda\sim10^{-13}$  GeV, which is much smaller than $\Gamma_\phi\sim10^{-7}$ GeV derived from the $\phi$-$h$ mixing. This implies an instantaneous reheating process after the FOPT, where $\phi$ directly rolls down to the true vacuum without undergoing oscillations. During this reheating, the temperature of the Universe increases monotonically from $T_1$ to $T_2$ over a very short period. Subsequently, DM particles are produced via freeze-in, resulting in an exponentially suppressed relic abundance $\Omega_Xh^2\propto e^{-2m_X/T_2}$. Therefore, we have a pure and vanilla WIMP freeze-in process without any significant dependence on the thermal history before the FOPT.

As a short summary, in the inflation-induced scenario, the reheating process is typically slow since $\Gamma_\varphi/H_I\ll1$. Consequently, the DM relic abundance not suppressed unless $X$ is a FIMP or $m_X$ is superheavy that its mass is related to the inflation model. In contrast, in our FOPT-induced scenario, it is natural to have an instantaneous reheating process as $\Gamma_\phi/H_\Lambda\gg1$. Therefore, the presence of the factor $e^{-2m_X/T_2}$ is guaranteed, allowing for the realization of WIMP freeze-in when $m_X\sim20\,T_2$. This distinction highlights the difference between the two scenarios. Our proposed scenario is novel due to its ability to easily and naturally achieve the desired exponential suppression factor while independent of the cosmic history before the FOPT.

\bibliographystyle{apsrev}
\bibliography{reference}

\end{document}